\documentclass[12pt,preprint]{aastex}

\usepackage{graphicx}
\usepackage{natbib}

\usepackage{amssymb}
\usepackage[figuresright]{rotating}
\newcommand{\lapprox}{$\stackrel {<}{_{\sim}}$}

\newcommand{\htwo}{H$_2$}

\newcommand{\li}{\ion{Li}{1}}

\newcommand{\sii}{[\ion{S}{2}]}
\newcommand{\hei}{\ion{He}{1}}
\newcommand{\caii}{\ion{Ca}{2}}
\newcommand{\oi}{[\ion{O}{1}]}

\newcommand{\um}{$\mu$m}

\newcommand{\msun}{M$_{\odot}$}
\newcommand{\rsun}{R$_{\odot}$}
\newcommand{\msunyr}{M$_{\odot}$\,yr$^{-1}$}
\newcommand{\macc}{$\dot{M}_{acc}$}

\newcommand{\mloss}{$\dot{M}_{loss}$}

\newcommand{\lacc}{$L_{\mathrm{acc}}$}
\newcommand{\rstar}{$R_{\mathrm{*}}$}

\newcommand{\mstar}{$M_{\mathrm{*}}$}

\slugcomment{To appear in Ap. J.}

\begin{document}

\title{A long-lasting quiescence phase of the eruptive variable V1118 Ori}
%\thanks{Based on observations collected at AZT-24 telescope
%(Campo Imperatore, Italy), AZT-8 (Crimea, Ukraine), and LX-200 (St.Petersburg, Russia)}}

\author{D.Lorenzetti\altaffilmark{1},
S.Antoniucci\altaffilmark{1},
T.Giannini\altaffilmark{1},
A. Harutyunyan\altaffilmark{2},
A.A.Arkharov\altaffilmark{3},
V.M.Larionov\altaffilmark{3,4,5},
F.Cusano\altaffilmark{6},
A.Di Paola\altaffilmark{1},
G.Li Causi\altaffilmark{1},
B.Nisini\altaffilmark{1},
R.Speziali\altaffilmark{1},
and
F.Vitali\altaffilmark{1}.
}
\altaffiltext{1}{INAF - Osservatorio Astronomico di Roma, via
Frascati 33, 00040 Monte Porzio, Italy,
dario.lorenzetti, simone.antoniucci, teresa.giannini, andrea.dipaola, gianluca.licausi, brunella nisini,
roberto.speziali, fabrizio.vitali@oa-roma.inaf.it}
\altaffiltext{2}{Fundaci\'{o}n Galileo Galilei – INAF, Telescopio Nazionale Galileo, 38700 Santa Cruz de la Palma, Tenerife, Spain, avet@tng.iac.es}
\altaffiltext{3}{Central Astronomical Observatory of Pulkovo,
Pulkovskoe shosse 65, 196140 St.Petersburg, Russia,
arkadi@arharov.ru}
\altaffiltext{4}{Astronomical Institute of St.Petersburg
University, Russia, vlar2@yandex.ru}
\altaffiltext{5}{Isaac Newton Institute of Chile, St.Petersburg branch}
\altaffiltext{6}{INAF - Osservatorio Astronomico di Bologna, via Ranzani 1, 40127 Bologna, Italy,
felice.cusano@oabo.inaf.it}
\begin{abstract}

V1118 Ori is an eruptive variable belonging to the EXor class of Pre-Main Sequence stars whose episodic outbursts
are attributed to disk accretion events. Since 2006, V1118 Ori is in the longest quiescence stage ever observed between two subsequent outbursts of its recent history. We present near-infrared photometry of V1118 Ori carried out during the last eight years, along with a complete spectroscopic coverage from 0.35 to 2.5 $\mu$m. A longterm sampling of V1118 Ori in quiescence has never been done, hence we can benefit from the current circumstance to determine the lowest values (i.e. the zeroes) of the parameters to be used as a reference for evaluating the physical changes typical of more active phases. A quiescence mass accretion rate between 1--3 $\times$ 10$^{-9}$ M$_{\sun}$ yr$^{-1}$ can be derived and the difference with previous determinations is discussed. From line emission and IR colors analysis a visual extinction of 1-2 mag is consistently derived, confirming that V1118 Ori (at least in quiescence) is a low-extinction T Tauri star with a bolometric luminosity of about 2.1 L$_{\sun}$. An anti-correlation exists between the equivalent width of the emission lines and the underlying continuum. We searched the literature for evaluating whether or not such a behaviour is a common 
feature of the whole class. The anti-correlation is clearly recognizable for all the available EXors in the optical range (H$\beta$ and H$\alpha$ lines), while it is not as much evident in the infrared (Pa$\beta$ and Br$\gamma$ lines).  The observed anti-correlation supports the accretion-driven mechanism as the most likely to account for continuum variations.

\end{abstract}


\section{Introduction}

After the accumulation of most of their final mass, young stellar objects (YSOs) of low-to-intermediate mass (0.5-8 $M_{\odot}$) appear as pre-main sequence objects while the mass accretion process continues at a lower rate. In this phase the source is expected to accrete from its circumstellar disk: matter moves through the viscous disk and eventually falls onto the star surface following the magnetic interconnection lines (Shu et al. 1994).
Observations show that the disk accretion process takes place through rapid and intermittent outbursts, usually detected at optical and near-IR wavelengths, which can be related to a sudden increase of the mass accretion rate by orders of magnitude (e.g. Hartmann \& Kenyon 1985, Antoniucci et al. 2008). An in-depth study of these accretion events is crucial to understand: ({\it i}) how the process eventually halts (thus determining the observed Initial Mass Function); ({\it ii}) how the bursts substantially alter the physical and chemical properties of the circumstellar disk, with major effects on the formation of proto-planetary systems; and ({\it iii}) the mechanism(s) at the origin of such outbursts.

Albeit a small and irregular photometric variations (typically 0.2-1 mag) caused by disk accretion
variability is a defining feature of all classical T Tauri stars (CTTSs), several young sources display powerful
outbursts of much larger intensity (up to 4-5 mag). Depending on different properties (burst duration, recurrence time between subsequent bursts, accretion rate, presence of absorption or emission lines), these objects are usually classified as FUors (Hartmann \& Kenyon 1985) and EXors (Herbig 1989). These latter are more suitable for an observational monitoring, since their variations typically occur on shorter times: outbursts duration from some months to one year superposed to longer quiescence periods (years); moreover, they present accretion rates of the order of (10$^{-6}$-10$^{-7}$ M$_{\sun}$ yr$^{-1}$), and emission line spectra (e.g. Lorenzetti et al. 2009, Sicilia-Aguilar et al. 2012, K\'{o}sp\'{a}l et al. 2011). The reader is also referred to the excellent review by Audard (2014) who gives a complete view of the EXor properties. 

Indeed, the nature of EXors is still very uncertain: no detailed analysis or modeling of 
their disk structure has been performed so far, so the mechanism responsible for the onset of the accretion
outbursts is not known. Proposed scenarios involve thermal instability in the disk or gravitational
instability triggered by a close companion (or planet). The latter alternative seems less probable, since in that case
subsequent outbursts should occur with some periodicity instead of repeating at irregular time intervals.

The very uncertain picture of EXor events stems not only from the small number of known EXor objects (around
two dozen, e.g. Lorenzetti et al. 2012; Audard et al 2014), but especially from the lack of a proper multi-wavelength monitoring of their photometric and spectroscopic properties, which prevented the construction of a comprehensive database of information for these objects. For this reason, we started an observational program dubbed EXORCISM (EXOR OptiCal and In frared Systematic Monitoring - Antoniucci et al. 2013) that is intended to perform a photometric and spectral monitoring in the range 0.5-2.5 $\mu$m of about 20 objects identified as known eruptive variables (EXor) or candidates.
In the framework of such a project, we present here optical and near-IR data of the typical source V1118 Ori collected during a quiescence phase in different dates from September 2006 to October 2014. 

During its recent history V1118 Ori underwent five documented outbursts, each lasting a couple of years (1982-84, 1988-90, 1992-94, 1997-98, 2004-06). Account for the first four events is given in Parsamian et al. 1993, Garc\'{i}a Garc\'{i}a \& Parsamian 2000, Herbig 2008 and references therein. For a complete view of the V1118 Ori properties shown during its last eruption, see Audard et al. 2005, 2010; Lorenzetti et al. 2006, 2007. Recently, Reipurth et al. (2007) discovered the binary nature of V1118 Ori, finding a close companion separated by 0.18 arcsec and about a factor 1.4 fainter in the H$\alpha$ filter ($\sim$ 0.4 mag). 

Typically, outburst phases of EXors are monitored more frequently than quiescence ones, although these latter have exactly the same relevance (e.g. Sipos et al. 2009). In fact, only during these phases the lowest values (the zeroes) of the different parameters can be accurately evaluated so allowing one to compute the physical changes, once the outburst values are obtained. Moreover, because of their longer duration, the quiescence phases can be observationally traced more easily than outbursting ones. 

Unfortunately, the quiescence status of V1118 Ori has never been thoroughly sampled, and in this paper we try
to fill this gap. In Sect.2 optical and near-IR observations are presented; the results are analysed 
in Sect.3, while our concluding remarks are given in Sect.4.
 
\section{Observations}

\subsection{Near-IR imaging}

\begin{figure}
\includegraphics[angle=0,width=18cm]{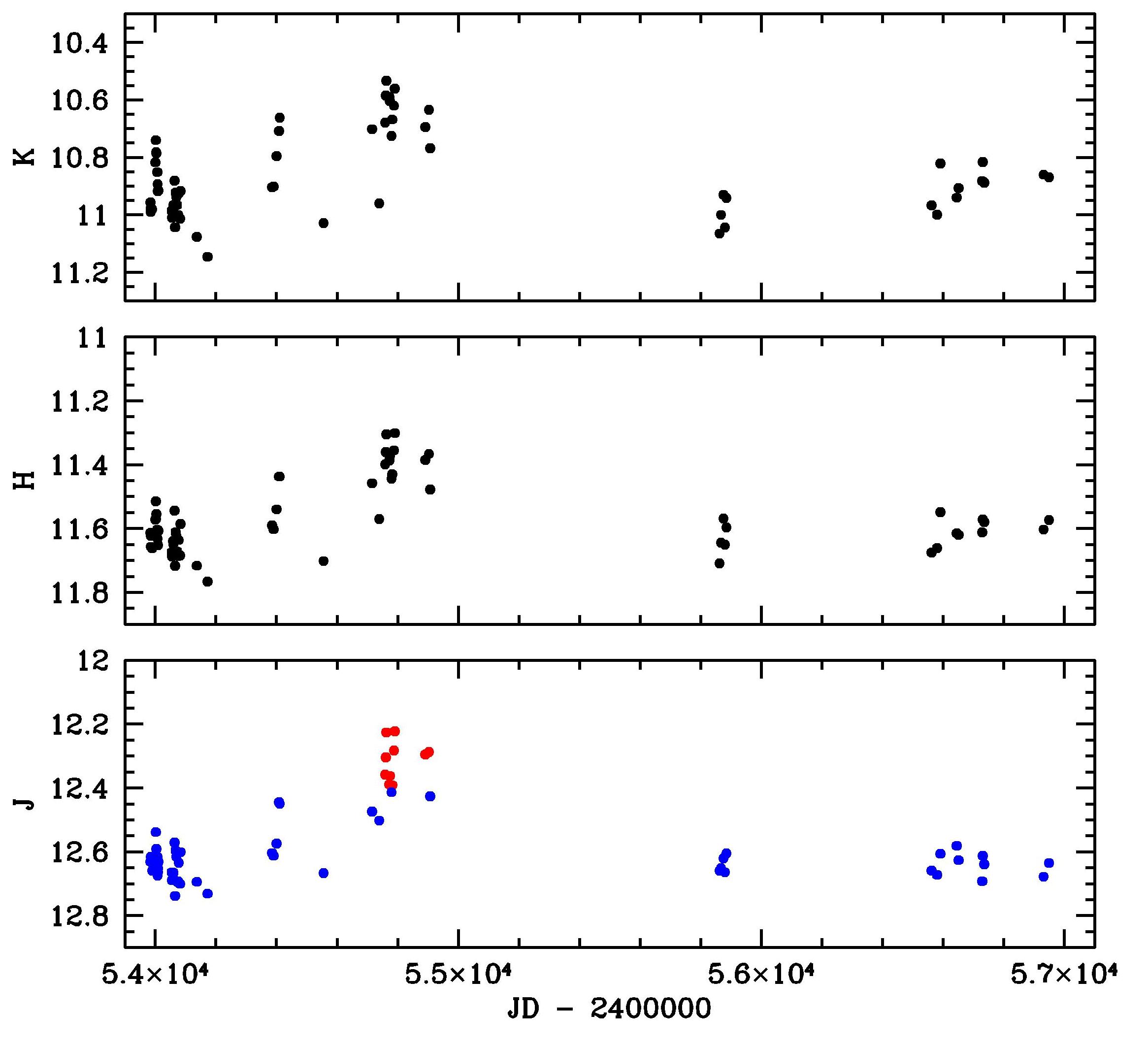}
   \caption{JHK light-curves of V1118 Ori. In the lower panel (J band) data points corresponding to the highest
   flux levels are depicted in red, while in blue those typical of a more quiescent state.
   \label{lightcurves:fig}}
\end{figure}

Near-IR data were obtained at the 1.1m AZT-24 telescope located at Campo Imperatore (L'Aquila - Italy) equipped with the
imager/spectrometer SWIRCAM (D'Alessio et al. 2000), which is based on a 256$\times$256 HgCdTe PICNIC array. Photometry was
performed with broad band filters J (1.25 $\mu$m), H (1.65 $\mu$m), and K (2.20 $\mu$m). 
%The total field of view is 4.4$\times$4.4 arcmin$^2$, which corresponds to a plate scale of 1.04 arcsec/pixel. 
All the observations were obtained by dithering the telescope around the pointed position. The raw imaging data were reduced by using standard procedures for bad pixel removal, flat fielding, and sky subtraction. Photometric data are listed in Table~\ref{nirphot:tab} while the derived light curves are depicted in the three panels of Figure~\ref{lightcurves:fig} for the J, H, and K band, respectively. In the bottom panel (J band) we evidence two groups of activity phases: 
a long-term quiescence state (blue dots) and a short period of moderate activity (red dots). These states have been arbitrarily identified with a J-band magnitude greater
(less) than 12.4, respectively. We will discuss these phases in deeper detail in Sect.3.1.

%Because of the seasonal observability of V1118 Ori, 
Near-IR data cover a period lasting eight years from September 2006 to October 2014; they have been collected without a systematic cadence and with no monitoring for long periods (of about one year), during which, however, no sign of outburst was reported (from both AAVSO circulars and Astronomer's Telegrams). Signs of modest activity (about 0.4 mag peak to peak) and short duration are superposed to such a long quiescence state (see Figure~\ref{lightcurves:fig}) that follows the last outburst, whose complete photometry is given elsewhere (Audard et al. 2010; Lorenzetti et al. 2006, 2007).\

%Complementary IR images were acquired on 2014 August 31 with NICS (Baffa et all. 2001) at the 4m Telescopio Nazionale %Galileo (TNG - La Palma, Canary Islands), by using the standard J, H, K broad-band filters and the H$_2$ narrow-band %filter centered at 2.12 $\mu$m. Here, the total field of view is roughly the same 4.2$\times$4.2 arcmin$^2$, but the %correspondingplate scale is 0.25 arcsec/pixel. The novelty of this set of images consists in a deep (about 1 hr) %integration on H$_2$ band, intended to firmly detect, or definitely rule out, any signpost of jet emission. 

\subsection{Optical spectroscopy}

\begin{figure}
\includegraphics[angle=0,width=17cm]{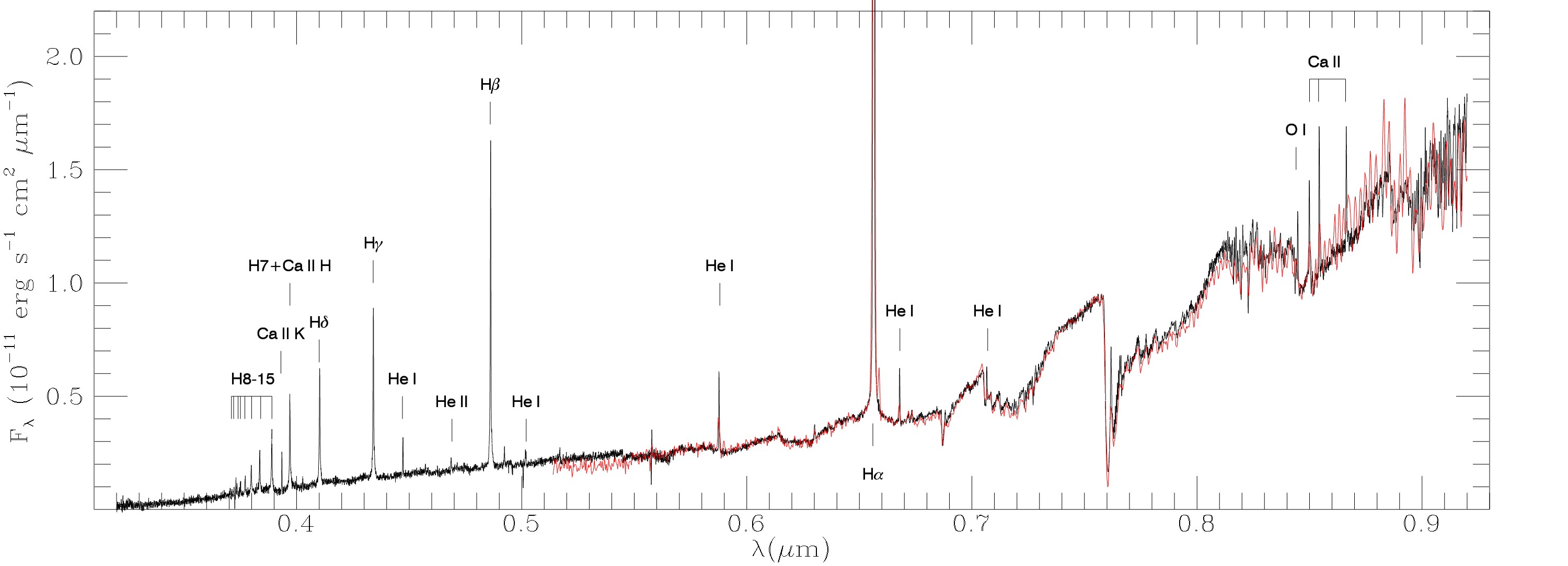}
   \caption{Optical spectra of V1118 Ori, taken with LBT/MODS (black) and TNG/DOLORES (red). Main emission features are indicated. 
   \label{opt_spec:fig}}
\end{figure}

Optical spectra were taken on two different occasions (2014 March 25 and August 12, JD 2\,456\,741 and 2\,456\,881, respectively). The first was obtained with the 8.4m Large Binocular Telescope (LBT) using the Multi-Object Double Spectrograph (MODS - Pogge et al. 2010).
The dual grating mode (Blue + Red channels) was used for a total integration time of 20 min to cover the 0.35 - 0.95 $\mu$m spectral range 
with a 0.6 arcsec slit (resolution $\sim$ 2000). The second spectrum was obtained with the 3.6 m Telescopio Nazionale Galileo (TNG) using the Device Optimized for the LOw RESolution (DOLORES) instrument. The low resolution red (LR-R) grism was used for a total integration time of 30 min to cover the 0.50-0.95 $\mu$m spectral range with a resolution of $\sim$ 700. In both cases (LBT and TNG) images were bias and flat-field corrected using standard procedures. 
After removing sky background, the two-dimensional spectra were extracted and collapsed to one dimension. For
both spectra, wavelength calibration was achieved through available lamp exposures, while spectral calibration was obtained from observations of spectro-photometric standards. The optical spectra of V1118 Ori are depicted in different colors (black for LBT, red for TNG) in Figure~\ref{opt_spec:fig}.
These are the only quiescence spectra at a high level of sensitivity comparable with that  of Herbig (2008) data. Both spectra show an extraordinary degree of repeatability concerning the continuum shape, whereas the derived line fluxes present variations of less than 20\%. Such an occurrence might be the result of the extreme steadiness of this quiescence state, that, as such, can be considered very suitable for deriving the parameters of the object in its low state.

\subsection{Near-IR spectroscopy} 

\begin{figure}
\includegraphics[angle=0,width=17cm]{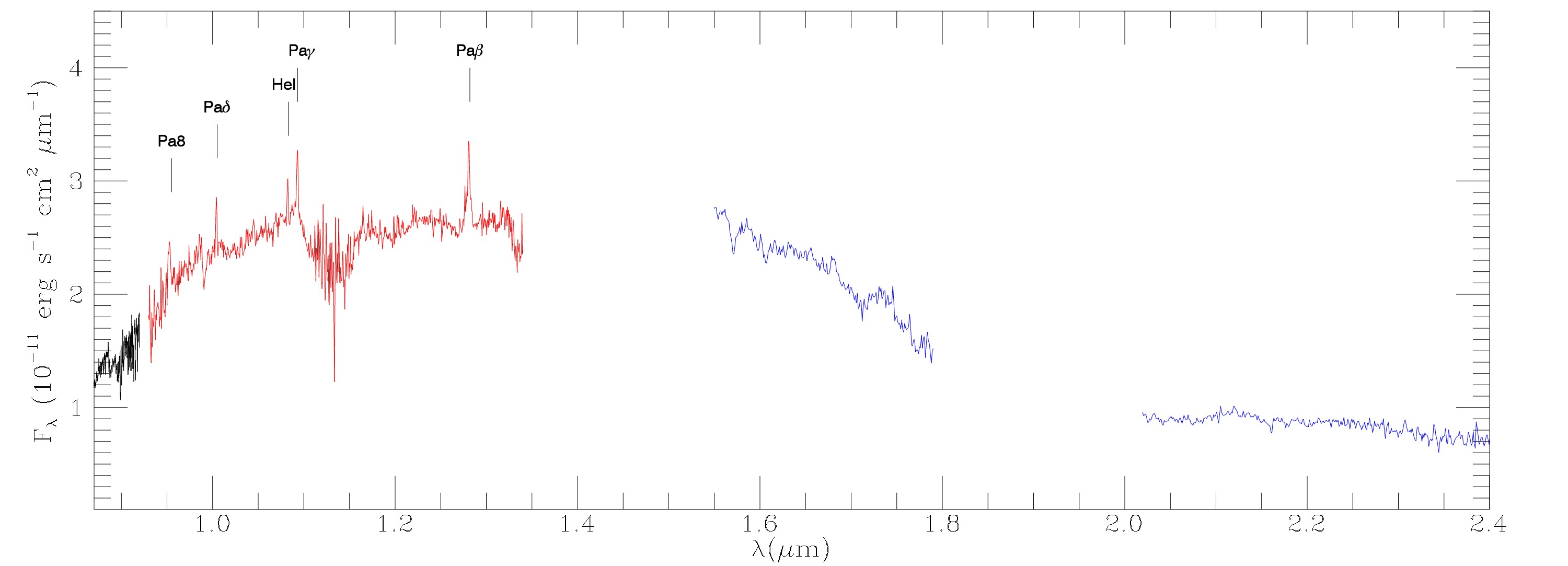}
   \caption{Near-IR spectrum of V1118 Ori, taken with TNG/NICS IJ (red) and HK (blue) grisms; spectral segments 
   that were heavily corrupted by telluric absorptions were removed. Main emission features are indicated.
   \label{nir_spec:fig}}
\end{figure}

A low resolution spectrum ($\mathcal{R}$ $\sim$ 500, slit width 1 arcsec) was obtained on 2014 August 26 
with NICS at TNG with two IR grisms IJ (0.90 -
1.45 $\mu$m) and HK (1.40 - 2.50 $\mu$m),in two subsequent exposures of 20 and 25 min, respectively. 
The standard ABB$\arcmin$A$\arcmin$ mode was exploited within a long slit oriented at a position angle of 40$^\circ$
in order to minimize the flux of a close field star entering the slit. The spectral images were flat-fielded, sky-subtracted, and corrected for the optical distortion in both
the spatial and spectral directions. Telluric features were removed by dividing the extracted spectra by that of a normalized telluric standard star, once corrected for its intrinsic spectral features. Wavelength and flux calibration were obtained from arc lamps and from our photometric data taken in the same period, respectively.
In Figure~\ref{nir_spec:fig} both portions (IJ and HK) of the resulting near-IR spectrum are given, labelling the relevant features identified.

\section{Results and analysis}

%As anticipated, V1118 Ori is currently in the longest quiescence status occurred in the last three decades. Therefore, the %quiescence parameters derived in this paper are expected to be enough reliable, since refer to a stable and long-lasting %period.

\subsection{Near-IR photometry}

Our images are not able to resolve the companion and no other high-angular resolution observations exist in other bands, 
hence we cannot precisely evaluate how the companion properties affect the quiescence parameters of V1118 Ori. Therefore, until this case will not be solved by future observations, we will not perform any (highly hypothetical) correction to the observed values derived for V1118 Ori, also considering that if a difference of 0.4 mag maintains over the entire spectral range, such correction would be marginal. 

\begin{figure}
\includegraphics[angle=0,width=18cm]{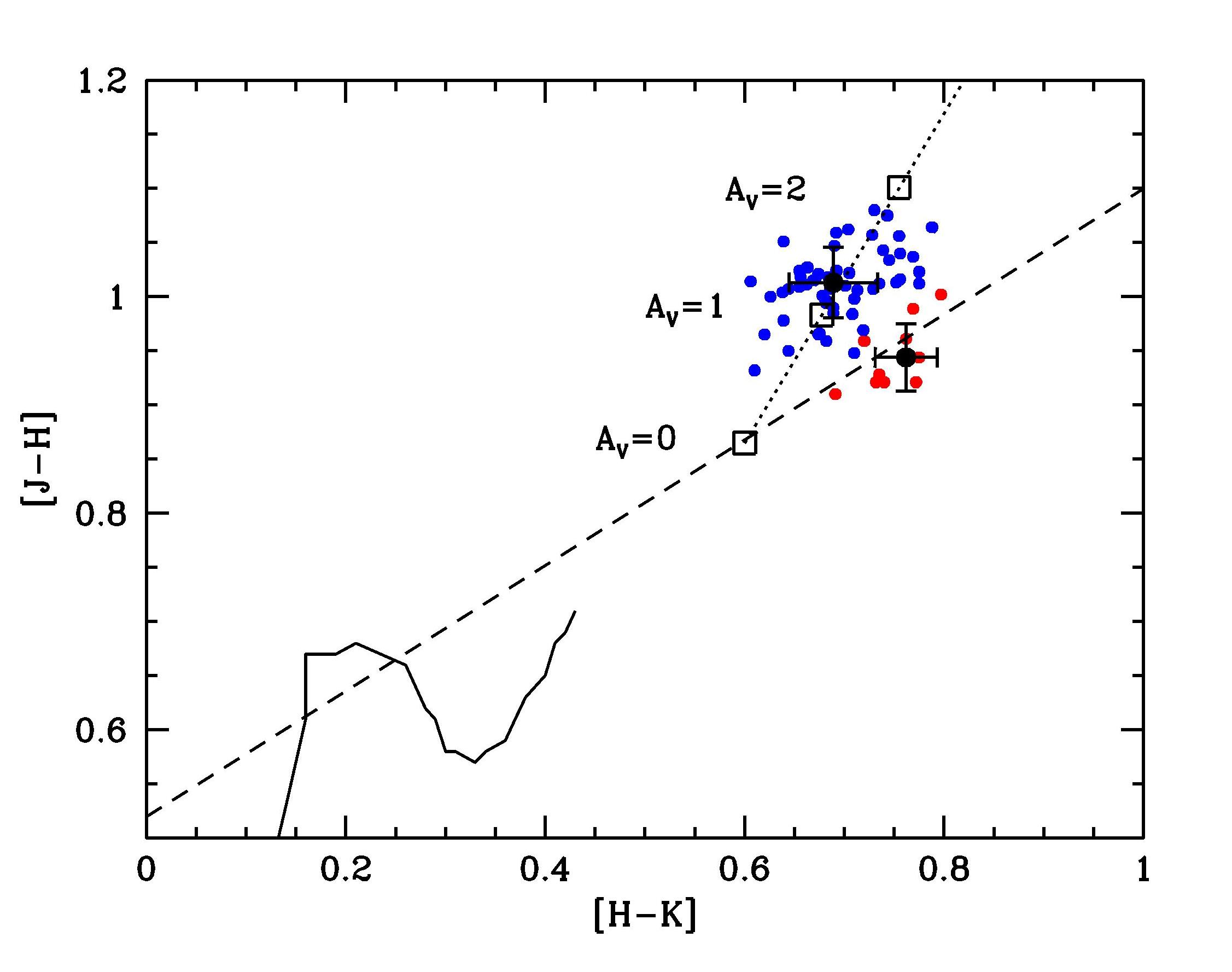}
  \caption{Near-IR two colors plot ([J-H] vs. [H-K]) of the quiescence period of V1118 Ori. 
Data points are those given in Table~\ref{nirphot:tab}. 
A separation between the color of the quiescence phase (blue - the same as in Figure~\ref{lightcurves:fig}) 
and that of the moderate activity (red) is recognizable. Black solid circles indicate the median values
of the two distribution (considering 1$\sigma$ error). The solid line marks the unreddened main sequence, 
whereas the dashed one is the locus of the T Tauri stars (Meyer et al. 1997). Black
dotted line represents the reddening law by Cardelli et al. (1989), where different values of 
A$_V$ are indicated by open squares. 
%The cross in the lower right corner indicates the typical error.
   \label{nircol:fig}}
\end{figure}

In Figure~\ref{nircol:fig} the near-IR two-colors plot is presented based on the photometric data given in 
Table~\ref{nirphot:tab}. Data points are depicted with different colors (blue and red) to indicate the two different
levels of activity evidenced in Figure~\ref{lightcurves:fig} (see Sect.2.1). From the inspection of the two-colors plot, two considerations can be retrieved. First, all 
the data roughly cluster in the locus typical of T Tauri stars of a late spectral type and very low extinction. 
In particular, a value of A$_V$ $\simeq$ 1-2 mag can reasonably account for the data distribution. 
Second, a separation between the color of the quiescence phase (blue) and that of the moderate activity occurred during years 2008-2009 (points in red) is recognizable. In particular, we depicted with black filled dots the median values
of the two distributions (considering 1$\sigma$ error). The same trend (i.e. that different colors are associated even with modest continuum  variations) has been found analysing also the light-curves of different EXors (Giannini et al. in preparation). 
%these suggest that the most
%significant variation occurred in the [J-H] color, i.e. the most sensitive to spot temperature variations.
Indeed, more significant decreases occurring in both the colors (i.e. a blueing effect) are commonly
associated to the outburst phases of EXors (Lorenzetti et al. 2007, Audard et al. 2010; see e.g. Fig.1 
of Lorenzetti et al. 2012). In this latter paper it is
discussed how these variations cannot be accounted for simply by extinction, but rather by an additional
thermal component of the emission. Such a circumstance seems to be confirmed also for the moderate
activity presented here, since the red points are located fairly orthogonal to the extinction curve rather
than along it.\

Taking into account the lowest photometric values even detected, Audard et al. (2010) provided a plot of the quiescence Spectral Energy Distribution (SED) (their Figure 15-left) and an estimate of 2.0 L$_\sun$ for the bolometric luminosity.
By substituting their JHK photometry with our values taken on 2007 Mar 12, and by applying the on-line fitting procedure\footnote{Available at http://caravan.astro.wisc.edu/protostars/sedfitter.php} (Robitaille et al. 2006, 2007),
we obtain practically the same value of bolometric luminosity (2.1 L$_\sun$). Moreover, the output parameters
(such as T$_\star$ = 4080 K and  M$_\star$ = 0.7 M$_\sun$) are well compatible with the known physical parameters of V1118 Ori (T$_\star$ $\simeq$ 3700 K;  M$_\star$ = 0.4 M$_\sun$ - Hillenbrand 1997).

\subsection{Optical and near-IR spectroscopy}

%The flux-calibrated spectra of V1118 Ori are displayed in Figs.~\ref{opt_spec:fig} and \ref{nir_spec:fig}.

The emission lines detected in the spectrum of the target are among those commonly observed in active young sources. A list of these lines is given in Table~\ref{lines:tab}.
The most prominent emission features are the HI recombination lines of the Balmer and Paschen series, which are commonly associated with flows of accreting gas (e.g. Calvet et al. 2000, 2004). Balmer lines are clearly visible up to H15 in the MODS spectrum, but no significant Balmer jump can be spotted at the end of the series, suggesting a fairly weak Balmer continuum.
The spectrum shows also many permitted transitions of \hei\ in the optical range and the \hei\ line at 1.08~\um, which is indicative of stellar winds (e.g. Edwards et al. 2006).
Other permitted lines are those of \caii\ (H and K line doublet at 0.39 $\mu$m and the triplet at 0.85-0.87\um) and the OI line at 0.845~\um.

Weak forbidden emission lines of \oi\ at 0.630 $\mu$m and \sii\ at 0.636 $\mu$m (which are typical tracers of shocks driven by jets, see e.g. Giannini et al. 2008) are not detected in the spectrum of the source, although they are present as
strong nebular contribution eliminated by the sky emission subtraction. This circumstance and the non-detection of [Fe~II] (e.g. lines at 0.71-0.74, 1.25, and 1.64~\um) and \htwo\ (2.12~\um) transitions indicate that, at least in the current quiescence state, there is no significant outflow activity from the source. This result is supported by the H$_2$ images
of the Catalogue of Molecular Hydrogen Emission-Line Objects (MHOs - Davis et al. 2010), where some \htwo\ filamentary structures are present in the field, but unrelated with V1118 Ori.

\begin{figure}
\includegraphics[angle=0,width=17cm]{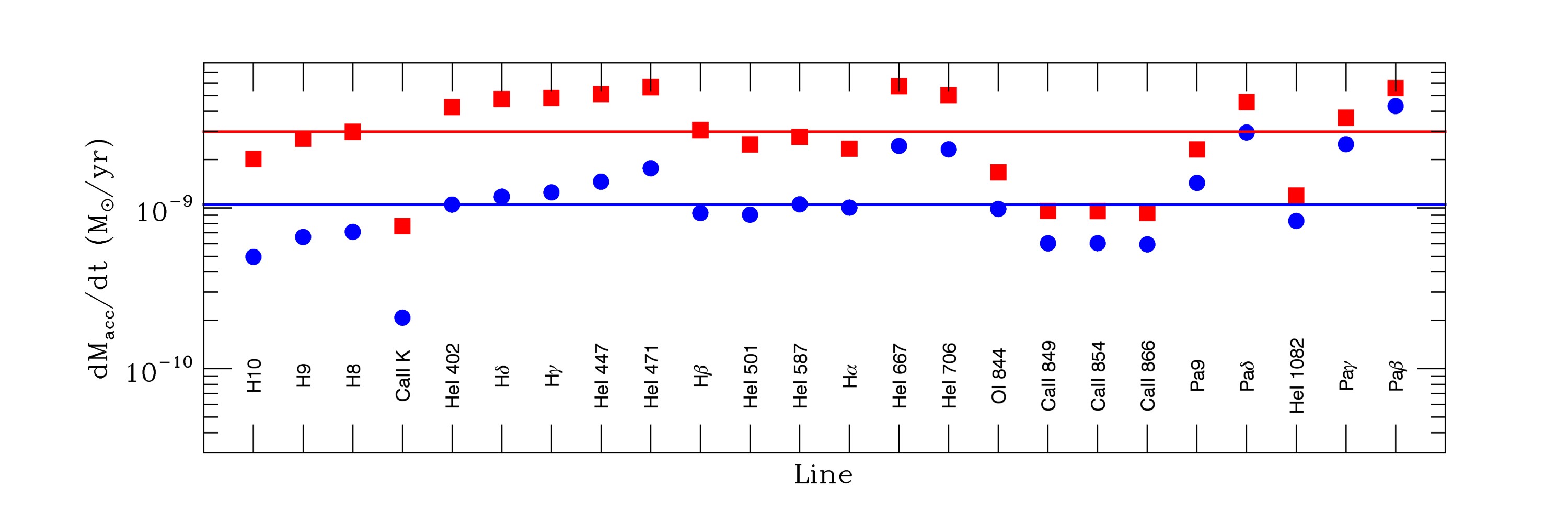}
   \caption{Mass accretion rate determinations computed from the flux of the indicated tracers using Alcal\'a et al. (2014) relationships, by assuming a visual extinction $A_V=1$ (blue circles) and $A_V=2$ (red squares). The median \macc\ values are marked for both cases with a solid line.
   \label{macc:fig}}
\end{figure}

From the observed emission lines we derived an estimate of the mass accretion rate (\macc) of the quiescent phase of V1118 Ori. For that we employed
the set of empirical relationships that connect line and accretion luminosity (\lacc) 
derived by Alcal\'a et. al (2014) in a sample of young active T Tauri stars in Lupus.
By adopting a distance of 400 pc, 
%and a visual extinction \av\ = 1 mag, 
we computed the accretion luminosity from 24 different tracers, avoiding blended lines. 
%and using the parametrisation by Cardelli et al. (1989) for dereddening the fluxes. 
The \lacc\ values thus inferred were then converted to mass accretion rates by using the relationship (e.g. Gullbring et al. 1998):
\begin{equation}
\label{eq:macc}
\dot{M}_{acc} = \frac{L_{acc}R_{*}}{GM_{*}} \left(1-\frac{R_{*}}{R_{in}}\right)^{-1} ~,
\end{equation}
where we assumed \mstar\ = 0.4\msun, \rstar\ = 1.29\rsun\ (Hillenbrand 1997, Stassun et al. 1999), and a typical inner radius $R_{in}=5 R_{*}$. The \macc\ values that we derive by adopting a visual extinction $A_V=1$ are reported in Table~\ref{lines:tab}. These are comprised in the interval 2 $\times 10^{-10}$ and 4 $\times 10^{-9}$, with a median \macc\ of about 1 $\times 10^{-9}$ \msunyr and a 1$\sigma$ dispersion of 7 $\times 10^{-10}$ \msunyr. By assuming $A_V=2$ instead, we obtain a median accretion rate of about 3 $\times 10^{-9}$ \msunyr with a dispersion of 1.6 $\times 10^{-9}$ \msunyr.
The single accretion rates derived in both cases from all 24 lines are shown in Fig.~\ref{macc:fig}.

%By using all lines, we have also computed the extinction value which minimizes the relative dispersion of the our 24 \macc\ estimates, finding $A_V$ = 2.2 mag. With this extinction value the median accretion rate becomes 3.9 $\times 10^{-9}$ \msunyr.

We note that our \macc\ estimates are much lower than the previously estimated quiescence \macc\ of 2.5 $\times 10^{-7}$\msunyr\ that was derived by Audard et al. (2010) from SED modelling; a further discussion on this aspect is given in Sect. 3.3.

%We note that an extinction of \av=3.3 mag can be estimated from the ratio Pa$\delta$/Br$\gamma$, 
%assuming that the emission is optically thin \citep[e.g.][]{bary08} and adopting Cardelli's law. 
%This yields a slightly lower median \macc\ value of 2$\times 10^{-8}$\msunyr.

\subsection{Comparison with the outburst phase}

%Once the quiescence parameters of V1118 Ori are stated, it is worthwhile to perform a first comparison with
%some values derived during more active phases. Such a comparison is just a preliminary attempt to estimate the gross %differences between the two states; indeed, a more comprehensive comparison should be made taking into account that some %parameters vary significantly over the course of the outburst itself.\
We perform here a comparison between our inferred quiescence parameters and those derived during more active phases.\

{\bf Bolometric luminosity L$_{bol}$} - From the SED fitting, a value of 2.0 L$_{\sun}$ is given (Audard et al. 2010 and
confirmed in Sect.3.1) for the quiescent luminosity. Different estimates are given for the outburst luminosity: 25.4 L$_{\sun}$ (Lorenzetti et al. 2006) and 7.4 L$_{\sun}$ (Audard et al. 2010). These values refer to two different outbursts that likely exhibit different levels of brightness, nevertheless a substantial increase of the luminosity certainly occurs, related to the appearance of the spot onto the stellar surface and to the thermal disk contribution.\

{\bf Visual extinction A$_V$} - From the near IR two-colors plot (Figure~\ref{nircol:fig}) and from line emission analysis, A$_V$ values of about 1-2 mag are consistently obtained. The extinction does not seem to change in the outburst during which Audard et al. (2010) derived, from X-rays column density, A$_V$ values of 1.7$^{+0.8}_{-0.6}$ mag or 1.4$^{+0.6}_{-0.5}$ mag, for R$_V$ equal to 3.1 or 5.5, respectively. A constant extinction suggests that changes of other physical parameters are not attributable to significant amounts of intervening dust (see also Audard et al. 2010; Lorenzetti et al. 2009).\

{\bf Mass accretion rate $\dot{M}$} - In Sect. 3.2 \macc\ values of 1 or 3 $\times 10^{-9}$ \msunyr\ were derived, depending on the assumed  A$_V$, 1 or 2 mag, respectively. This value is two orders of magnitude lower than the previously estimated quiescence \macc\ of  2.5 $\times 10^{-7}$\msunyr\ derived by Audard et al. (2010) from SED modelling.
This discrepancy is likely related to the different methods employed for the computation of the accretion rate. Accretion estimates from the emission lines are to be regarded as more reliable, since the lines are believed to trace the accretion columns or the strong accretion-related winds from the object.
Moreover, the empirical accretion luminosity-line luminosity relationships, which are widely used as a proxy for deriving  \macc, are directly calibrated from measurements of the UV excess emission from the accretion shock (e.g. Alcal\'{a} et al. 2014, Muzerolle et al. 1998).
Finally, we note that the \macc\ values we find for the quiescence of V1118 Ori are completely consistent with the accretion rates measured in most T Tauri objects (e.g. Natta et al. 2006, Alcal\'{a} et al. 2008, Biazzo et al. 2012, Antoniucci et al. 2011, 2014).

For a comparison, we applied the emission line method to compute \macc\ to our 2006 data (note that at that time the
relationships for deriving \macc\ from line fluxes were not available yet). From three bright Pashen lines (Pa$\beta$,
Pa$\gamma$, and Pa$\delta$) we get \macc\ value $\sim$ 1.0 $\times 10^{-7}$\msunyr\, namely around 2 orders of magnitude
higher than the quiscence value. Notably, this value is lower not only than the outburst value of
Audard et al. (2010) of  1.0 $\times 10^{-6}$\msunyr\, but even than their quiescence value of 2.5 $\times 10^{-7}$\msunyr\ .
While the difference between the two outburst values could be ascribed to the different outburst phases
sampled in Lorenzetti et al. (2006) and in Audard et al. (2010), the inconsistency between our outburst value and the
quiescence one by Audard et al. can only be a result of the different computation methods. Finally, we note that our outburst determination of \macc\ / \mloss\ is within the range (roughly 2-20) predicted by jet launching models (Shu et al. 2000, K\"onigl \& Pudritz 2000, Ferreira, Dougados \& Cabrit, 2006), if we take the value of \mloss\ = 4.0 $\times 10^{-8}$\msunyr\ obtained from the  HI recombination lines.\

{\bf Emission lines} - As listed in Table~\ref{lines:tab} the optical-IR quiescence spectrum of V1118 Ori is dominated by HI and \hei\ recombination lines. Very few and faint metallic lines are also present. Such a situation is
the same reported by Herbig (2008), who observed the source twice, one in outburst and one in a fading phase. While in the former phase he noticed the prominence of H~I, He~I, Ca~II, Fe~I, Fe~II,
%\hei\ , \caii\ , \fe\ , \feii\ , 
and other neutral and ionized metals, when the star was much fainter the spectrum had changed radically: only HI and \hei\ emission lines remained strong. Also our near-IR spectrum, taken during the last outburst (Lorenzetti et al. 2006) showed metallic (e.g. Na~I) and CO emission features,
now completely disappeared, even if we observed at higher sensitivity. In particular, CO emission is usually associated with large amounts of warm gas (T$\sim$3000 K) in the inner regions of the circumstellar disc, which is indicative of an active phase of accretion (Najita et al. 1996; Lorenzetti et al. 2009; K\'{o}sp\'{a}l et al. 2011).
During a similar quiescence phase, Herbig (2008) observed also TiO absorption bands and the \li\ 6707 absorption line. While TiO bands are clearly observed in our spectrum, \li\ is absent, possibly because of a superposition of emission and absorption, both recurrent on different activity stages of V1118 Ori (Herbig 2008). However, we should have been able to detect a faint \li\ feature similar to that observed by Herbig at just a 3$\sigma$ level ($\simeq$ 7 $\times 10^{-17}$
erg s$^{-1}$ cm$^{-2}$) over the continuum. 
%We can say that our upper limit at that wavelength is 2 $\times$ 10$^{-16}$ erg s$^{-1}$ cm$^{-2}$), hence we should have 
%been able to detect any \li\ feature similar to that observed by Herbig. 
In our quiescence spectrum Ca~II emission lines are also present, but no comparison can be made with the Herbig (2008) spectrum that did not cover this spectral range. However, the presence of Ca~II lines cannot be considered a peculiarity, as they are quite ubiquitous in T Tauri systems.\

% In the upper left panel of Figure~\ref{EW:fig}
%the relation between the H$\beta$ EW and the B band continuum is depicted. A clear (anti-)correlation exists, which will %be discussed in the next section, where this kind of analysis is expanded to include other EXor system and other spectral %bands to evaluate whether or not the EW behaviour of V1118 Ori can be considered to be a general property of the class.

\subsection{Equivalent widths and continuum}

In the general context of the debate on whether the EXor events are accretion or extinction driven, it is worthwhile to analyse the relationship between lines and continuum emission. In particular, we aim at comparing the line emission behaviour in phases characterized by a different level of activity,  i.e. how the EW varies with the increasing underlying continuum. 

First of all we notice that the only member of the EXor class analyzed in this respect (PVCep) shows a clear correlation (typical regression coefficient $\geq$ 0.9) between line and continuum flux (Lorenzetti et al. 2013). Since HI-recombination lines are considered as good tracers of the accretion rate (e.g. Muzerolle et al. 1998) the observed correlation supports the idea that accretion-driven mechanism is the most likely to account for continuum variations. Notice that accretion or mass loss processes are both compatible with the observed behaviour, since their ratio \macc\ / \mloss\ is roughly in the range
2-20. Although line and continuum fluxes are correlated, we can provide a more quantitative information by directly comparing the line EW and the continuum itself.

\begin{figure}
\includegraphics[angle=0,width=18cm]{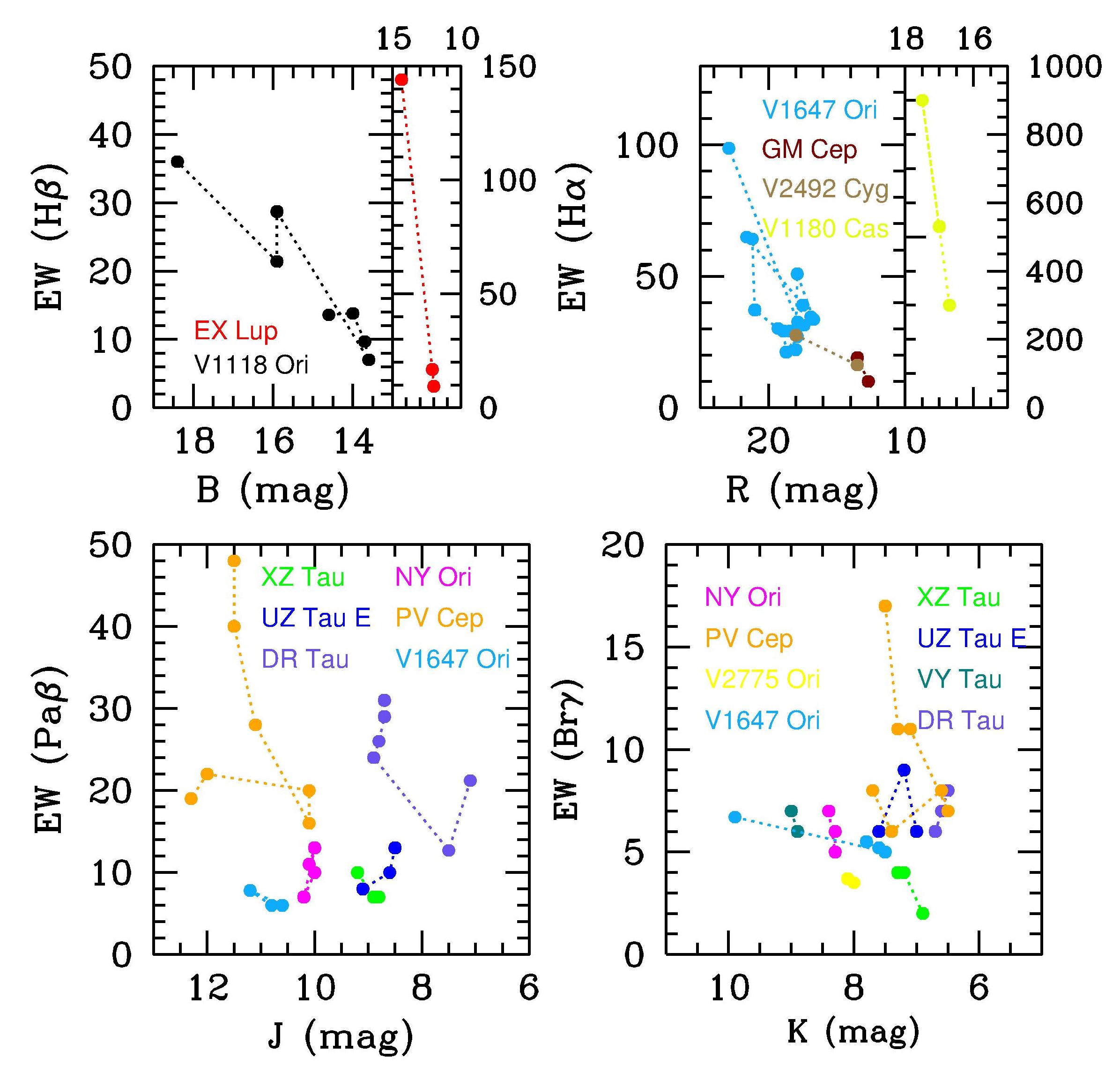}
   \caption{EW of optical (upper panels) and IR (lower panels) HI recombination lines as a function of the underlying continuum expressed in magnitudes. To avoid a possible confusion, the EW values are labelled (in the y-axis) with a positive sign, although they are conventionally negative for emission lines, as reported in Tables~\ref{EW1:tab} and \ref{EW2:tab}.
   \label{EW:fig}}
\end{figure}

Literature data concerning the EW variations of prominent emission lines and those of their underlying continua are taken starting from the list of EXors and candidates  (Lorenzetti et al. 2012, their Table 1). 
We report in Tables~\ref{EW1:tab} and \ref{EW2:tab} only the data that refer to simultaneous observations of HI recombination lines H$\beta$ and H$\alpha$ and the continuum in the bands B and R, respectively. 
Providing that intra-day or day time-scale variations are usually expected to be modest, in few occasions, the values of EW and continuum have been assigned to the same date (within a maximum distance of 3-4 days), even if not strictly simultaneous. 
%Our selection process does not take into account those cases in which no variation (within the errors) occurs %simultaneously in both parameters. 
In Figure~\ref{EW:fig} the results of our literature search are depicted.
The same exercise was done for the near-IR recombination lines Pa$\beta$, Br$\gamma$ and
the relative photometry in the J, K bands, respectively. For the sake of compactness, near-IR data are not listed in a Table, but are also presented in the same Figure~\ref{EW:fig} and discussed below.\

The anti-correlation, evident for V1118, occurs also for the majority of other sources when the optical lines 
(H$\beta$ and H$\alpha$) are considered, while for the near-IR lines (Pa$\beta$ and Br$\gamma$) the anti-correlation is not as much evident, being ascertainable only for about 50\% of the presented cases.
However, if we take into account only the most significant variations, namely those with $\Delta$EW $\geq$ 100\%, or $\Delta$mag $\geq$ 2, the total percentage (considering both optical and near-IR lines) of anti-correlations increases to about 80\%. In conclusion, the existence of an anti-correlation can be reasonably confirmed as a general property of the EXor class of objects: a result already supported by previous investigations on individual objects (e.g. Cohen et al. 1981, Magakian \& Movsessian 2001, Acosta-Pulido et al. 2007).

The anti-correlation indicates that the continuum presents larger (likely faster) variations than those of the
lines, which means that continuum and lines do not obey to a common mechanism of heating and cooling. 
%Such fluctuations are likely due to the {\it direct} radiation from the hot spot and an
%{\it indirect} brightening due to the radiation from the innermost regions of the circumstellar disk. The combined
%effect of such double component with the time-lag needed for the radiation propagation, 
%could originate the observed anti-correlation. 
Such a circumstance tends to rule out the variable extinction as
cause of the observed variability, since in that case a constant value
of EW should be expected for any continuum fluctuation. Even a selective obscuration can be ruled out: 
dust should be located at the dust condensation zone, very close to the star and, as such, it should 
obscure the stellar photosphere more than the accretion columns and the wind regions.

\section{Concluding remarks}

We have presented the results of a monitoring of the EXor system V1118  Ori. Data concerning both IR photometry and optical/near IR spectroscopy cover a long lasting period of 8 years (2006-2014). This period starts from the last outburst
and is the longest state of quiescence ever monitored. Hence it is very suitable for deriving reliable parameters
typical of this status, that can be used as a reference to evaluate physical changes occurring during more active phases.

\begin{itemize}
\item[-] Near IR colors of V1118 Ori cluster in a locus typical of a late type T Tauri star embedded in a low-extinction
(A$_V$ $\simeq$ 1-2 mag) environment. A trend of the colors to become bluer when the source flux increases, can be recognized even in presence of modest ($\simeq$ 0.4 mag) variations.
\item[-] From model fitting, stellar parameters (T$_\star$ = 3900 K;  M$_\star$ = 0.7 M$_\sun$) well compatible with the known physical parameters of V1118 Ori are derived and the value of 2.0 L$_{\sun}$ for the bolometric luminosity in quiescence is confirmed.
\item[-] Optical/near IR spectra are dominated by both H~I recombination lines and \hei\ permitted transitions. TiO absorption bands are also identified, without any sign of forbidden transitions, indicating that (at least in the current quiescence state) there is no significant outflow activity from the source.
\item[-] Emission line fluxes are consistently exploited to derive a mass accretion rate ($\dot{M}$) of
1--3 $\times$ 10$^{-9}$ M$_{\sun}$ yr$^{-1}$ compatible with typical values observed in T Tauri stars. This value is substantially different from that previously derived using a different
method. This discrepancy is likely related to the different methods employed for the computation of the accretion rate. Accretion estimates from the emission lines are to be regarded as more reliable, since the lines are believed to trace the accretion columns or the strong accretion-related winds from the object.
\item[-] An (anti-)correlation exists between the equivalent width of the emission lines and the underlying continuum. A search in the literature confirms that the behaviour of V1118 Ori is a common 
feature of the whole class and supports the accretion-driven mechanism as the most likely to account for continuum variations.
\end{itemize}

In Table~\ref{parameters:tab} we summarize the results presented above: this table
can be a useful reference for delineating the V1118 Ori properties at quiescence.

\section{Acknowledgements}

Based on observations made with different instrument. [1] the Italian Telescope Galileo (TNG) operated on the island of La Palma by the Fundaci\'{o}n Galileo Galilei of the INAF (Istituto Nazionale di Astrofisica) at the Spanish Observatorio del Roque
de los Muchachos of the Istituto de Astrofisica de Canarias; [2] the Large Binocular Telescope (LBT). The LBT is an international collaboration among institutions in the United States, Italy and Germany. LBT Corporation partners are: The University of Arizona on behalf of the Arizona university system; Istituto Nazionale di Astrofisica, Italy; LBT Beteiligungsgesellschaft, Germany, representing the Max-Planck Society, the Astrophysical Institute Potsdam, and Heidelberg University; The Ohio State University, and The Research Corporation, on behalf of The University of Notre Dame, University of Minnesota and University of Virginia. [3] the AZT-24 IR Telescope at Campo Imperatore (L'Aquila - Italy) operated under the
responsability of the INAF-Osservatorio Astronomico di Roma (OAR).

{}

\newpage

%%%%%%%%%%%%%%%%%%%%%%%%%%%%%%%%%%%%%%%%%%%%%%%%%%%%%%%%%%%%%%%%%%%%%%%
%%%%%%%%%%%%%%%%%%%%%%%%%%%%%%%%%%%%%%%%%%%%%%%%%%%%%%%%%%%%%%%%%%%%%%%

%%%%%%%%%%%%%%%%%%%%%%%%%%%%%%%%%%%%%%%%%%%%%%%%%%%%%%%%%%%%%%%%%%%%%%%
%%%%%%%%%%%%%%%%%%%%%%%%%%%%%%%%%%%%%%%%%%%%%%%%%%%%%%%%%%%%%%%%%%%%%%%
%
%%%%%%%%%%%%%%%%%%%%%%%%%%%%%%%%%%%%%%%%%%%%%%%%%%%%%%%%%%%%%%%%%%%%%%%
%%%%%%%%%%%%%%%%%%%%%%%%%%%%%%%%%%%%%%%%%%%%%%%%%%%%%%%%%%%%%%%%%%%%%%%

%%%%%%%%%%%%%%%%%%%%%%%%%%%%%%%%%%%%%%%%%%%%%%%%%%%%%%%%%%%%%%%%%%%%%%%
%%%%%%%%%%%%%%%%%%%%%%%%%%%%%%%%%%%%%%%%%%%%%%%%%%%%%%%%%%%%%%%%%%%%%%%

%%%%%%%%%%%%%%%%%%%%%%%%%%%%%%%%%%%%%%%%%%%%%%%%%%%%%%%%%%%%%%%%%%%%%%%
%\begin{figure}
%\includegraphics[angle=0,width=18cm]{fit_rob.jpg}
%  \caption{Best fit 'a la Robitaille' to the lowest photometric values exhibited by V1118 Ori. 
%   \label{rob:fig}}
%\end{figure}
%%%%%%%%%%%%%%%%%%%%%%%%%%%%%%%%%%%%%%%%%%%%%%%%%%%%%%%%%%%%%%%%%%%%%%%

%%%%%%%%%%%%%%%%%%%%%%%%%%%%%%%%%%%%%%%%%%%%%%%%%%%%%%%%%%%%%%
%\begin{figure}
%\includegraphics[angle=0,width=17cm]{opt_spec.jpg}\\
%\includegraphics[angle=0,width=17cm]{nir_spec.jpg}
%   \caption{\textit{Top}: optical spectra of V1118 Ori, taken with LBT/MODS (black) and TNG/DOLORES (red). 
%   \textit{Bottom}: Near-IR spectrum of V1118 Ori, taken with TNG/NICS IJ (red) and HK (blue) grisms; spectral segments 
%   that were heavily corrupted by telluric absorptions were removed. 
%   Main emission features are indicated.
%   \label{spec:fig}}
%\end{figure}
%%%%%%%%%%%%%%%%%%%%%%%%%%%%%%%%%%%%%%%%%%%%%%%%%%%%%%%%%%%%%%%%%%%%%%%%

%%%%%%%%%%%%%%%%%%%%%%%%%%%%%%%%%%%%%%%%%%%%%%%%%%%%%%%%%%%%%%%%%%%%%%%
%%%%%%%%%%%%%%%%%%%%%%%%%%%%%%%%%%%%%%%%%%%%%%%%%%%%%%%%%%%%%%%%%%%%%%%

%%%%%%%%%%%%%%%%%%%%%%%%%%%%%%%%%%%%%%%%%%%%%%%%%%%%%%%%%%%%%%%%%%%%%%%
%%%%%%%%%%%%%%%%%%%%%%%%%%%%%%%%%%%%%%%%%%%%%%%%%%%%%%%%%%%%%%%%%%%%%%%

%%%%%%%%%%%%%%%%%%%%%%%%%%%%%%%%%%%%% TABLE 1: NIRPHOT %%%%%%%%%%%%%%%%%%%%%%%%%%%%%%%%%%%%%%%%%%%%%%%
\begin{deluxetable}{ccccc|ccccc}
\tabletypesize{\footnotesize} \tablecaption{Near-IR photometry of V1118 Ori. \label{nirphot:tab}} 
\tablewidth{0pt}
\tablehead{Date      &  JD     &  J    &   H   &   K      & Date      &  JD     &  J    &   H   &   K      \\
         (yy/mm/dd)  &+2400000 & \multicolumn{3}{c}{(mag)}& (yy/mm/dd)&+2400000 & \multicolumn{3}{c}{(mag)} }
\startdata   
%     05 Oct 26       & 53669.6        & 11.47 & 10.55 &  9.70        
%     05 Nov 03       & 53677.6        & 11.16 & 10.34 &  9.47        
%     05 Nov 04       & 53678.5        & 11.28 & 10.43 &  9.52        
%     05 Nov 08       & 53682.6        & 10.74 & 10.01 &  9.21  
%     05 Nov 09       & 53683.6        & 11.00 & 10.20 &  9.40  
%     05 Nov 10       & 53684.5        & 10.96 & 10.17 &  9.36  
     06 Sep 06       & 53984.6        & 12.63 & 11.61 & 10.96  &  07 Nov 05 & 54410.6 & 12.45 & 11.44 & 10.66   \\ 
     06 Sep 07       & 53985.6        & 12.62 & 11.66 & 10.97  &  08 Mar 29 & 54555.3 & 12.67 & 11.70 & 11.03   \\
     06 Sep 11       & 53989.6        & 12.66 & 11.66 & 10.98  &  08 Sep 06 & 54715.6 & 12.47 & 11.46 & 10.70   \\
     06 Sep 22       & 54000.6        & 12.63 & 11.57 & 10.82  &  08 Sep 29 & 54738.6 & 12.50 & 11.57 & 10.96   \\                                                                               
     06 Sep 23       & 54001.6        & 12.54 & 11.51 & 10.74  &  08 Oct 19 & 54758.6 & 12.36 & 11.40 & 10.68   \\
     06 Sep 24       & 54002.6        & 12.63 & 11.57 & 10.78  &  08 Oct 21 & 54760.6 & 12.30 & 11.36 & 10.58   \\
     06 Sep 25       & 54003.6        & 12.59 & 11.55 & 10.78  &  08 Oct 23 & 54762.6 & 12.23 & 11.30 & 10.53   \\
     06 Sep 28       & 54006.6        & 12.62 & 11.60 & 10.85  &  08 Nov 02 & 54772.6 & 12.39 & 11.39 & 10.59   \\
     06 Sep 29       & 54007.6        & 12.67 & 11.63 & 10.89  &  08 Nov 03 & 54773.6 & 12.36 & 11.37 & 10.60   \\
     06 Sep 30       & 54008.6        & 12.66 & 11.65 & 10.92  &  08 Nov 09 & 54779.5 & 12.41 & 11.44 & 10.72   \\
     06 Oct 01       & 54009.6        & 12.65 & 11.60 & 10.91  &  08 Nov 11 & 54781.5 & 12.39 & 11.43 & 10.67   \\
     06 Oct 02       & 54010.6        & 12.63 & 11.61 & 10.91  &  08 Nov 17 & 54787.5 & 12.28 & 11.35 & 10.62   \\
     06 Nov 15       & 54054.5        & 12.66 & 11.67 & 10.98  &  08 Nov 20 & 54790.5 & 12.21 & 11.30 & 10.56   \\
     06 Nov 16       & 54055.6        & 12.69 & 11.69 & 11.01  &  09 Mar 01 & 54891.2 & 12.30 & 11.38 & 10.69   \\  
     06 Nov 19       & 54058.5        & 12.66 & 11.64 & 10.98  &  09 Mar 13 & 54903.3 & 12.29 & 11.37 & 10.63   \\
     06 Nov 20       & 54059.5        & 12.67 & 11.65 & 10.97  &  09 Mar 17 & 54907.3 & 12.43 & 11.48 & 10.77   \\
     06 Nov 24       & 54063.5        & 12.57 & 11.54 & 10.88  &  11 Oct 27 & 55861.6 & 12.66 & 11.71 & 11.06   \\ 
     06 Nov 26       & 54065.5        & 12.74 & 11.72 & 11.04  &  11 Nov 01 & 55866.6 & 12.65 & 11.64 & 11.00   \\
     06 Nov 28       & 54067.6        & 12.60 & 11.61 & 10.92  &  11 Nov 09 & 55874.6 & 12.62 & 11.57 & 10.93   \\ 
     06 Nov 30       & 54069.5        & 12.61 & 11.62 & 10.94  &  11 Nov 14 & 55879.5 & 12.66 & 11.65 & 11.04   \\
     06 Dec 02       & 54071.5        & 12.69 & 11.67 & 10.97  &  11 Nov 19 & 55884.5 & 12.60 & 11.60 & 10.94   \\
     06 Dec 06       & 54075.5        & 12.69 & 11.69 & 11.00  &  13 Sep 26 & 56561.6 & 12.66 & 11.67 & 10.97   \\
     06 Dec 08       & 54077.4        & 12.63 & 11.64 & 10.93  &  13 Oct 14 & 56579.6 & 12.67 & 11.66 & 11.00   \\  
     06 Dec 12       & 54081.5        & 12.70 & 11.68 & 11.01  &  13 Oct 25 & 56590.6 & 12.61 & 11.55 & 10.82   \\ 
     06 Dec 14       & 54083.5        & 12.60 & 11.59 & 10.92  &  13 Dec 18 & 56644.5 & 12.58 & 11.61 & 10.94   \\  
     07 Feb 05       & 54137.3        & 12.69 & 11.72 & 11.08  &  13 Dec 23 & 56650.4 & 12.62 & 11.62 & 10.91   \\                            
     07 Mar 12       & 54172.3        & 12.73 & 11.77 & 11.15  &  14 Mar 11 & 56728.3 & 12.69 & 11.61 & 10.88   \\
     07 Oct 11       & 54385.6        & 12.60 & 11.59 & 10.90  &  14 Mar 13 & 56730.3 & 12.61 & 11.57 & 10.82   \\
     07 Oct 16       & 54390.6        & 12.61 & 11.60 & 10.90  &  14 Mar 18 & 56735.3 & 12.64 & 11.58 & 10.89   \\   
     07 Oct 26       & 54400.6        & 12.57 & 11.54 & 10.79  &  14 Sep 30 & 56930.7 & 12.68 & 11.60 & 10.86   \\   
     07 Nov 03       & 54408.6        & 12.44 & 11.44 & 10.71  &  14 Oct 18 & 56948.6 & 12.63 & 11.57 & 10.87   \\           
\enddata
\tablenotetext{a}{Typical errors of the near-IR magnitudes do not exceed 0.03 mag}
%\tablecomments{~~}
\end{deluxetable}

\begin{deluxetable}{lccc}
\tabletypesize{\footnotesize} 
\tablecaption{Detected emission lines, unreddened fluxes, and derived accretion rates assuming a visual extinction A$_V$ = 1 mag. \label{lines:tab}} 
\tablewidth{0pt}
\tablehead{Line ID  &  $\lambda$   &   Flux $\pm \Delta$ Flux            & \macc    \\
                    & ($\mu$m)     &(10$^{-16}$ erg s$^{-1}$ cm$^{-2}$)  & (\msunyr) }
\startdata  
\multicolumn{4}{c}{{\bf LBT/MODS}}                               \\
\hline
     H15       & 0.371      &   0.9   $\pm$  0.3   &  --        \\
     H14       & 0.372      &   0.9   $\pm$  0.3   &  --        \\
     H13       & 0.374      &   0.9   $\pm$  0.3   &  --        \\
     H12       & 0.375      &   1.6   $\pm$  0.3   &  --        \\
     H11       & 0.377      &   3.2   $\pm$  0.4   &  --        \\
     H10       & 0.380      &   5.0   $\pm$  0.4   &  5.0E-10   \\
     H9        & 0.384      &   8.4   $\pm$  0.5   &  6.6E-10   \\
     H8        & 0.389      &  12.4   $\pm$  0.5   &  7.1E-10   \\
   Ca II K     & 0.393      &   4.5   $\pm$  0.2   &  2.1E-10   \\
H7 + Ca II H   & 0.397      &  23.9   $\pm$  0.5   &  --        \\
  He I         & 0.403      &   1.7   $\pm$  0.3   &  1.0E-09   \\
     H$\delta$ & 0.410      &  28.4   $\pm$  0.4   &  1.2E-09   \\
     H$\gamma$ & 0.434      &  41.7   $\pm$  0.4   &  1.3E-09   \\
  He I         & 0.447      &   4.4   $\pm$  0.2   &  1.5E-09   \\
  Ti II        & 0.457      &   1.4   $\pm$  0.3   &  --        \\
  He II        & 0.469      &   2.1   $\pm$  0.3   &  --        \\
  He I         & 0.471      &   0.9   $\pm$  0.2   &  1.8E-09   \\
     H$\beta$  & 0.486      &  66.6   $\pm$  0.4   &  9.3E-10   \\
  He I         & 0.501      &   1.7   $\pm$  0.2   &  9.1E-10   \\
  ?            & 0.502      &   1.0   $\pm$  0.2   &  --        \\
  ?            & 0.517      &   1.8   $\pm$  0.4   &  --        \\
  Mg I         & 0.518      &   0.8   $\pm$  0.2   &  --        \\
  He I         & 0.588      &   9.5   $\pm$  0.4   &  1.1E-09   \\
  \oi$^{a}$    & 0.630      &   3.2   $\pm$  0.4   &  --        \\
     H$\alpha$ & 0.656      & 567.2   $\pm$  0.5   &  1.0E-09   \\
  He I         & 0.668      &   8.8   $\pm$  0.8   &  2.4E-09   \\
  \sii$^{a}$   & 0.671      &   3.1   $\pm$  0.8   &  --        \\
  \sii$^{a}$   & 0.673      &   3.4   $\pm$  0.7   &  --        \\
  He I         & 0.707      &   6.7   $\pm$  2.3   &  2.3E-09   \\
  He I         & 0.837      &   7.2   $\pm$  1.2   &  --        \\
  O I          & 0.845      &  15.7   $\pm$  2.7   &  9.9E-10   \\
   Ca II       & 0.850      &  22.5   $\pm$  1.0   &  6.0E-10   \\
   Ca II       & 0.854      &  24.6   $\pm$  0.7   &  6.0E-10   \\
   Ca II       & 0.866      &  20.7   $\pm$  0.8   &  5.9E-10   \\
  He I         & 0.873      &   4.8   $\pm$  1.4   &  --        \\
    Pa12       & 0.875      &  12.0   $\pm$  2.2   &  --        \\ 
  He I         & 0.886      &   8.1   $\pm$  1.1   &  --        \\
    Pa9        & 0.923      &  24.0   $\pm$  8.5   &  1.4E-09   \\
\hline                                                          
\multicolumn{4}{c}{\bf{TNG/DOLORES}} \\                         
\hline                                                                                                                  
  He I         & 0.588      &  12.7   $\pm$  1.5   &  1.5E-09   \\
%  \oi$^{a}$    & 0.630      &   5.6   $\pm$  1.2   &  --        \\
     H$\alpha$ & 0.656      & 656.8   $\pm$  1.9   &  1.2E-09   \\
  He I         & 0.668      &   8.2   $\pm$  1.5   &  2.2E-09   \\
%  \sii$^{a}$   & 0.671      &   8.6   $\pm$  2.8   &  --        \\
%  \sii$^{a}$   & 0.673      &   7.2   $\pm$  2.3   &  --        \\
\hline                                                          
\multicolumn{4}{c}{\bf{TNG/NICS}} \\                            
\hline                                                                                                                     
    Pa11       & 0.882      &   53.0  $\pm$ 11.0   &  --        \\                   
    Pa9        & 0.923      &   62.5  $\pm$ 10.0   &  --        \\
    Pa$\delta$ & 1.004      &   85.1  $\pm$  8.2   &  2.9E-09   \\
  He I         & 1.082      &   68.5  $\pm$  6.3   &  8.3E-10   \\
    Pa$\gamma$ & 1.092      &   105.7 $\pm$ 10.0   &  2.5E-09   \\
    Pa$\beta$  & 1.281      &   262.3 $\pm$ 21.4   &  4.3E-09   \\
\enddata
%\tablenotetext{a}{~~nebular}
%\tablenotetext{b}{~~Values of I$_c$ magnitude estimated from a light-curve plot}
%\tablenotetext{c}{~~Values of r$^{\prime}$ magnitude estimated from a light-curve plot}
%\tablenotetext{d}{~~Values of R$_c$ magnitude}
%\tablecomments{The first four lines...}

\end{deluxetable}  

%%%%%%%%%%%%%%%%%%%%%%%%%%%%%%%%%%%%% TABLE 2: EW (H_beta) %%%%%%%%%%%%%%%%%%%%%%%%%%%%%%%%%%%%%%%%%%%%%%%
\begin{deluxetable}{cccccc}
\tabletypesize{\footnotesize} \tablecaption{EW of the H$\beta$ emission line vs. B band underlying continuum measured in different epochs. \label{EW1:tab}} 
\tablewidth{0pt}
\tablehead{Source  &  Date       &  JD      & EW     &  B     &   Ref. \\
                   & (yy/mm/dd)  &+2400000  &(\AA)   & (mag)  &         }
\startdata   
V1118 Ori          & 89 Jan 07   & 47533.5  &- 13.6  &   14.6 &   1    \\
                   & 89 Jan 11   & 47537.5  &- 13.8  &   14.0 &   1    \\
                   & 89 Feb 01   & 47558.5  &- 9.7   &   13.7 &   1    \\
                   & 89 Dec 19   & 47879.5  &- 7.0   &   13.6 &   1    \\
                   & 92 Dec 02   & 48958.5  &- 28.7  &15.6-16.2 &   2  \\
                   & 93 Feb 22   & 49040.5  &- 21.4  &15.6-16.2 &   2  \\
                   & 14 Mar 25   & 48958.5  &- 36.0  &   18.4 &   3    \\
                   &             &          &        &        &        \\
EX Lup             &  93 Apr     &   ...    &- 24.0  &   13.2 &   4    \\
                   &  94 Mar 03  & 49414.5  &- 9.4   &   12.0 &   4    \\
                   &  07 Jul     &   ...    &- 144   &   14.4 &   5    \\
\enddata
\tablenotetext{-}{~~Errors on EW and magnitude are typically $\sim$0.2 \AA~ and \lapprox 0.1 mag, respectively}
%\tablenotetext{b}{~~Values of I$_c$ magnitude estimated from a light-curve plot}
%\tablenotetext{c}{~~Values of r$^{\prime}$ magnitude estimated from a light-curve plot}
%\tablenotetext{d}{~~Values of R$_c$ magnitude}

\tablecomments{References to the Tables from 2 and 3: (1) Parsamian et al. 1996; (2) Parsamian et al. 2002; (3) this paper; (4) Herbig et al. 2001 (and reference therein): (5) Sipos et al. 2009; (6) Kun et al. 2011; (7) Aspin \& Reipurth 2009; (8) Ojha et al. 2006; (9) Aspin et al. 2009; (10) Aspin et al. 2011b: (11) Aspin et al. 2011a; (12) Sicilia-Aguilar et al. 2008; %(13) Lorenzetti et al. 2009; (14) Brittain et al. 2010.
}
\end{deluxetable}

%%%%%%%%%%%%%%%%%%%%%%%%%%%%%%%%%%%%% TABLE 3: EW (H_alpha) %%%%%%%%%%%%%%%%%%%%%%%%%%%%%%%%%%%%%%%%%%%%%%%
\begin{deluxetable}{cccccc}
\tabletypesize{\footnotesize} \tablecaption{EW of the H$\alpha$ emission line vs. R band underlying continuum measured in different epochs. \label{EW2:tab}} 
\tablewidth{0pt}
\tablehead{Source  &  Date       &  JD      & H$\alpha$ EW   &  B        &   Ref. \\
                   & (yy/mm/dd)  &+2400000  &(\AA)           & (mag)     &         }
\startdata  
V1180 Cas          &  03 Feb 05  & 52675.5  &- 300           &16.7~$^a$  &   6    \\  
                   &  08 Aug 28  & 54706.5  &- 900           &17.5~$^a$  &   6    \\ 
                   &  10 Dec 31  & 55561.5  &- 530           &17.0~$^a$  &   6    \\ 
                   &             &          &                &           &        \\ 
V1647 Ori          &  04 Feb 14  & 53049.5  &- 31.4          &17.4~$^b$  &   7    \\                                
                   &  04 Feb 22  & 53057.5  &- 34.6          &16.9       &   8    \\
                   &  04 Feb 23  & 53058.5  &- 33.6          &16.7       &   8    \\                                
                   &  04 Mar 10  & 53074.5  &- 50.9          &17.9~$^b$  &   7    \\
                   &  04 Oct 06  & 53284.5  &- 29.5          &18.0~$^b$  &   7    \\
                   &  04 Nov 16  & 53325.5  &- 22.1          &18.0~$^b$  &   7    \\
                   &  04 Dec 12  & 53351.5  &- 29.2          &18.4~$^b$  &   7    \\
                   &  05 Jan 08  & 53378.5  &- 27.1          &17.9~$^b$  &   7    \\
                   &  05 Aug 30  & 53612.5  &- 21.2          &18.7~$^b$  &   7    \\
                   &  05 Sep 25  & 53638.5  &- 29.2          &18.9~$^b$  &   7    \\
                   &  05 Oct 13  & 53656.5  &- 30.2          &19.3~$^b$  &   7    \\
                   &  05 Nov 18  & 53692.5  &- 37.2          &21.0~$^b$  &   7    \\
                   &  05 Nov 27  & 53701.5  &- 64.2          &21.2~$^b$  &   7    \\
                   &  05 Dec 25  & 53729.5  &- 64.9          &21.6~$^b$  &   7    \\
                   &  06 Jan 05  & 53740.5  &- 66.6          &21.9~$^b$  &   7    \\
                   &  06 Feb 16  & 53782.5  &- 98.7          &22.9~$^b$  &   7    \\
                   &  08 Aug 31  & 54694.5  &- 39            &17.5~$^b$  &   9    \\
                   &  11 Feb 02  & 55594.5  &- 32.6          &17.8~$^b$  &   10   \\
                   &             &          &                &           &        \\ 
V2492 Cyg          &  10 Sep 25  & 55464.5  &- 16.3          &  13.5     &   11   \\                                                       
                   &  10 Nov 05  & 55505.5  &- 27.6          &  18.0     &   11   \\                                                       
                   &             &          &                &           &        \\
GM Cep             &  01 Jul 11  & 52101.5  &- 10            &  12.7     &   12   \\                                                       
                   &  07 Apr 27  & 54217.5  &- 19            &  13.5     &   12   \\
\enddata  
\tablenotetext{-}{~~Notes and references are the same as in Table~\ref{EW1:tab}} 
\tablenotetext{a}{~~Values of I$_c$ magnitude estimated from a light-curve plot}
\tablenotetext{b}{~~Values of r$^{\prime}$ magnitude estimated from a light-curve plot}
%\tablenotetext{d}{~~Values of R$_c$ magnitude}                      
\end{deluxetable}      

%\input{tab4.tex}
%\input{tab5.tex}
%%%%%%%%%%%%%%%%%%%%%%%%%%%%%%%%%%%%% TABLE 6: Quiescence parameters  %%%%%%%%%%%%%%%%%%%%%%%%%%%%%%%%%%%%%%%
\begin{deluxetable}{ccc}
\tabletypesize{\normalsize} \tablecaption{Quiescence parameters of V1118 Ori. \label{parameters:tab}} 
\tablewidth{0pt}
\tablehead{\multicolumn{2}{c}{Parameter}  &  Value     }
\startdata   
Distance                &   d          &    400 pc                            \\
Spectral Type           & SpT          &    M1e                               \\
Stellar radius          & R$_{*}$      &    1.29 R$_{\sun}$                   \\
Stellar mass            & M$_{*}$      &    0.41 M$_{\sun}$                   \\
\hline
Bolometric luminosity   & L$_{bol}$    &    2.1 L$_{\sun}$                   \\
Visual extinction       & A$_{V}$      &    1-2 mag                           \\
Mass accretion rate     & $\dot{M}$    &  1-3 10$^{-9}$ M$_{\sun}$ yr$^{-1}$  \\
\enddata
%\tablenotetext{-}{~~Errors on EW and magnitude are typically $\sim$0.2 \AA~ and \lapprox 0.1 mag, respectively}
%\tablenotetext{b}{~~Values of I$_c$ magnitude estimated from a light-curve plot}
%\tablenotetext{c}{~~Values of r$^{\prime}$ magnitude estimated from a light-curve plot}
%\tablenotetext{d}{~~Values of R$_c$ magnitude}

\tablecomments{The first four lines of the table list literature parameters, while the rest are derived in the present work.}
\end{deluxetable}

\end{document}